\documentclass[%
 reprint,
showpacs,preprintnumbers,
 amsmath,amssymb,
 aps,
]{revtex4-1}
\usepackage{graphicx}
\usepackage{dcolumn}
\usepackage{bm}
\usepackage{epstopdf}

\begin{document}

\preprint{APS/123-QED}

\title{Ultraviolet energy dependence of particle production sources in relativistic heavy-ion collisions}

\author{Georg Wolschin}%
 \email{g.wolschin@thphys.uni-heidelberg.de}
\affiliation{%
 Institut f{\"ur} Theoretische 
Physik
der Universit{\"a}t Heidelberg, Philosophenweg 16, D-69120 Heidelberg, Germany, EU\\
}%


\date{\today}

\begin{abstract}
The energy dependence of particle production sources in relativistic heavy-ion collisions is investigated from RHIC to LHC energies.
Whereas charged-hadron production in the fragmentation sources follows a $\ln(s_{NN}/s_0)$ law, particle production in the mid-rapidity gluon-gluon source exhibits a much stronger dependence $\propto \ln^3(s_{NN}/s_0)$, and becomes dominant between RHIC and LHC energies. The production of particles with pseudorapidities beyond the beam rapidity is also discussed.

\end{abstract}

\maketitle
\section{\label{sec:intro}Introduction\protect} 
The investigation of charged-hadron production in relativistic heavy-ion collisions has generated a vast amount of energy- and centrality-dependent data at energies reached at both, the Relativistic Heavy-Ion Collider RHIC \cite{alv11}, and the Large Hadron Collider LHC \cite{cole14}. It has been shown \cite{gw13} within the framework of a nonequilibrium-statistical relativistic diffusion model (RDM) \cite{gw99,gw04} that the energy-dependent multiplicity of produced charged hadrons is well understood quantitatively based on a mid-rapidity low-$x$ gluonic source and the two fragmentation sources. This applies not only to AuAu collisions at RHIC \cite{wob06} and PbPb at LHC \cite{rgw12}, but also to asymmetric systems such as dAu at RHIC \cite{wobi06} and $p$Pb at LHC \cite{gw13}.

The relativistic diffusion model is in scope and character located between the (equilibrium) statistical model for multiple hadron production that was proposed by Fermi \cite{fer50} and Hagedorn \cite{ha68}, and much more detailed numerical models that aim at a microscopic description of the collision, such as the Color Glass Condensate (CGC, see \cite{ge10}) for the initial state, hydrodynamics for the main part of the time evolution (e.g. \cite{koi07,luro08,alv10,hesne13}), and codes like URQMD for the final state \cite{bas13}. 

The statistical hadronization (or thermal) model has been further developed and compared to a large amount of data by many authors such as Braun-Munzinger et al. or Becattini et al. \cite{mabe08,pbm95,aa06}, and it has consistently -- with only few exceptions -- provided good descriptions of particle production yields, in particular, at mid rapidity. 
As a consequence of its ambition to account for particle production with few parameters (temperature, chemical potential, characteristic volume) in an equilibrium setting with collective expansion, the thermal model does, however, not describe effects such as the plateau occurring in rapidity distributions $dN/dy$ of produced particles at higher (RHIC and above) energies, the corresponding dip in pseudo rapidity $dN/d\eta$, and other outstanding features such as limiting fragmentation at RHIC and LHC energies. 


To account for such non-equilibrium effects and model the collision in full detail requires in current scenarios to match the CGC initial state smoothly to viscous hydrodynamics when the coupling constant becomes too strong in the course of the time development for perturbative QCD techniques to be applicable \cite{ga13}, and finally use Cooper-Frye freeze out \cite{cofr74} or another code that accounts for the final-state interactions \cite{bas13}. 

However, since even the most sophisticated codes that purport to describe the full time evolution will contain a certain amount of arbitrariness and can not fully replace the experiment, it appears indicated to permit phenomenological models such as the RDM that include non-equilibrium effects to some extent, reproduce substantial features of the data and have some predictive power, but do not claim to fully account for every detail of the collision and of the ensuing particle production. 

The nonequilibrium-statistical relativistic diffusion model is -- in its linear approximation \cite{gw04} -- based on an analytically solvable transport equation with three sources. It does not only consider particle production from a central source as the thermal model does, but also from the fragmentation sources. The latter evolve in time and eventually tend to merge with the central source towards an overall thermal equilibrium distribution, but since the interaction time is extremely short at RHIC and LHC energies, this equilibrium state is not reached, and in particular the rapidity and pseudorapidity distributions show characteristic nonequilibrium features. 

In this work I present an investigation of the energy dependence of the charged-hadron production sources within the relativistic diffusion model
in symmetric systems, AuAu at RHIC c.m. energies per nucleon pair of 19.6, 62.4, 130 and 200 GeV, and PbPb at LHC energies of 2.76 and 5.52 TeV. 
The gluon-dominated source, in addition to the fragmentation sources related to the valence part of the nucleons, had been implemented earlier into the RDM \cite{gw04,wob06}.
A related model with a gluonic source at mid rapidity had also been proposed by Bialas and Czyz \cite{bia05}. 

In \cite{gw13} it has been found that the fragmentation sources for produced charged hadrons -- which are clearly visible in net-proton rapidity distributions where the gluonic source cancels out \cite{mtw09} --  have the expected logarithmic dependence on $\sqrt{s_{NN}}$, whereas the
particle content in the mid rapidity gluon-gluon induced source that rises strongly with energy is close to a power law. This result has since been corroborated through other independent investigations of charged-particle and transverse energy production \cite{sah14,sami14} such that a renewed and more precise consideration in particular of the central source is indicated. 

The fragmentation sources are responsible for most of the yield in the regions close to the beam rapidities.
Here limiting fragmentation scaling \cite{alv11} is valid not only at RHIC, but also at LHC energies \cite{rgw12}. This is in contrast to earlier predictions of the thermal model \cite{cley08} which find a violation of extended longitudinal scaling at LHC energies, providing another indication that equilibrium statistical concepts are invalid in the fragmentation region. Here the yields in pseudorapidity also extend beyond the value of the beam rapidity, and in the final paragraph of this note the origin of this effect is discussed.

\section{Hadron production sources}

For a detailed phenomenological investigation of the charged-hadron particle content in the three particle-production sources, the nonequilibrium-statistical relativistic diffusion model \cite{gw99,gw04,gw13} is used. The fragmentation sources $R_{1,2}(y,t=\tau_{int})$ with charged-particle content $N_{ch}^{qg,1}$ (projectile-like), $N_{ch}^{gq,2}$ (target-like) and the midrapidity low-$x$ gluon-gluon source $R_{gg}(y,t = \tau_{int})$ with charged-particle content $N_{ch}^{gg}$
are added incoherently to generate the total pserudorapidity density distribution as
\begin{eqnarray}
    \lefteqn{
\frac{dN^{tot}_{ch}(y,t=\tau_{int})}{dy}=N_{ch}^{qg,1}R_{1}(y,\tau_{int})}\nonumber\\&&
\qquad\qquad +N_{ch}^{gq,2}R_{2}(y,\tau_{int})
+N_{ch}^{gg}R_{gg}(y,\tau_{int})
\label{normloc1}
\end{eqnarray}
with the rapidity $y = 0.5\cdot \ln((E+p)/(E-p))$, and the interaction time $\tau_{int}$. The latter corresponds to the total integration time of the underlying partial differential equation, which is a linear partial differential equation of the Fokker-Planck type, as described in \cite{gw13}. 

Converting the rapidity distribution $dN/dy$ for produced charged hadrons to the corresponding  pseudorapidity distribution $dN/d\eta$ ($\eta$ = - ln(tan($\theta/2$)) ) with the proper Jacobian transformation $dy/d\eta$ and minimizing the analytical solutions of the transport equation with respect to available pseudorapidity data then yields the particle content of the sources as functions of  $\sqrt{s_{NN}}$ \cite{gw13}. The corresponding RDM-parameters for central collisions have been published in Tab.~1 of \cite{gw13}.

\begin{figure}[tph]
\begin{center}
\includegraphics[width=8.8cm]{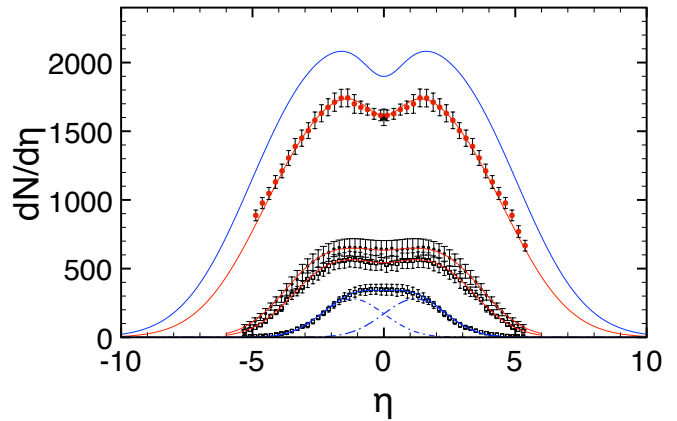}
\caption{\label{fig1}(Color online) The RDM pseudorapidity distribution functions for charged hadrons in central AuAu (RHIC) and PbPb (LHC) collisions
at c.m. energies of 19.6 GeV, 130 GeV, 200 GeV, 2.76 TeV and 5.02 TeV shown here are optimized in $\chi^2-$fits with respect to the PHOBOS \cite{bb03,alv11} (bottom) and ALICE \cite{gui13} (top) data, with parameters from \cite{gw13}. The upper distribution function is an extrapolation to the LHC design energy of 5.52 TeV. At the lowest energy, only the fragmentation sources contribute (dash-dotted curves).}

\end{center}
\end{figure}
\begin{figure}[tph]
\begin{center}
\includegraphics[width=8.4cm]{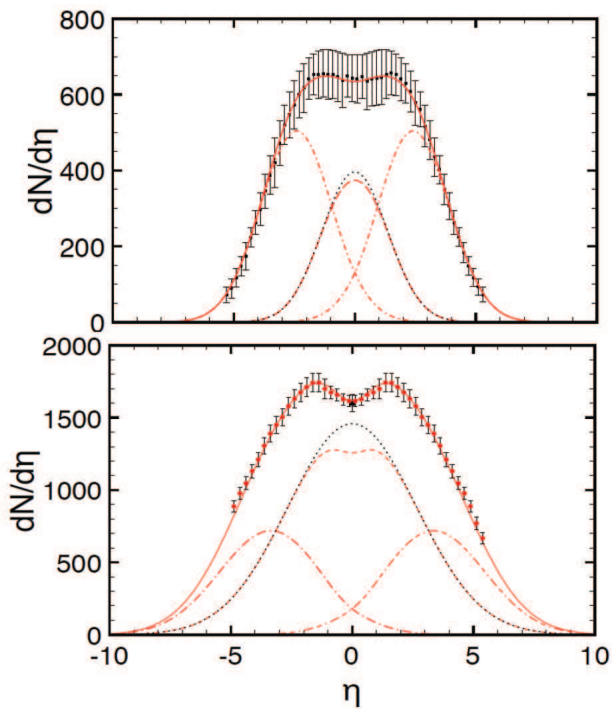}
\caption{\label{fig2}(Color online) The RDM pseudorapidity distribution functions for charged hadrons in central 200 GeV AuAu (top frame) and 2.76 TeV PbPb collisions
are adjusted through $\chi^2-$minimizations to the PHOBOS \cite{alv11} (see also \cite{wob06}) and ALICE  \cite{gui13} data, see \cite{gw13}. The underlying particle production sources are shown: dash-dotted curves are the fragmentation sources, dashed curves the mid rapidity gluon-gluon sources, and dotted curves the central sources without the effect of the Jacobian transformation from rapidity to pseudorapidity. The particle content in the gluon-gluon source rises strongly with increasing c.m. energy, and constitutes the largest source at LHC energies.}
\end{center}
\end{figure}
\begin{figure}[tph]
\begin{center}
\includegraphics[width=9cm]{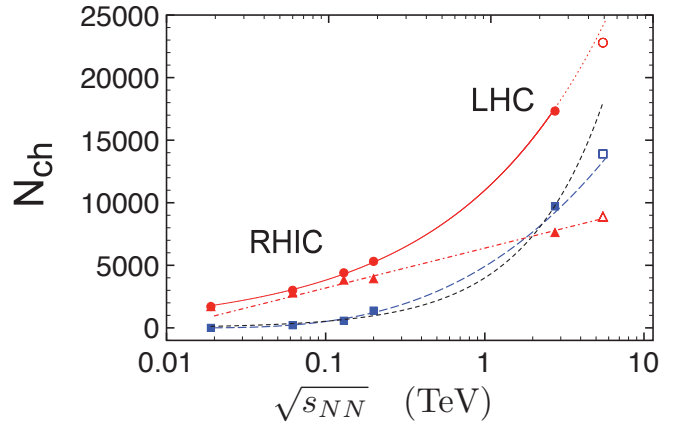}
\caption{\label{fig3}(Color online) Number of produced charged hadrons as function of the c.m. energy $\sqrt{s_{NN}}$ from RDM-fits of the available data for central heavy-ion collisions at 0.019, 0.062, 0.13, 0.2 TeV (RHIC, AuAu), 2.76 TeV (LHC, PbPb), plus extrapolation to 5.52 TeV. Circles are the total numbers, following a power law $\propto s_{NN}^{0.23}$. Triangles are particles from the fragmentation sources $\propto\log(s_{NN}/s_0)$. Squares are hadrons produced from the midrapidity source, with a dependence
 $\propto\log^3(s_{NN}/s_0).$  A power law $\propto s_{NN}^{0.44}$ \cite{gw13} is also shown (short-dashed curve), but fails to  fit the extrapolated 5.52 TeV yield.
The gluon-gluon source (dashed) becomes the main source of particle production between RHIC and LHC energies.}
\end{center}
\end{figure}
\begin{figure}[tph]
\begin{center}
\includegraphics[width=9cm]{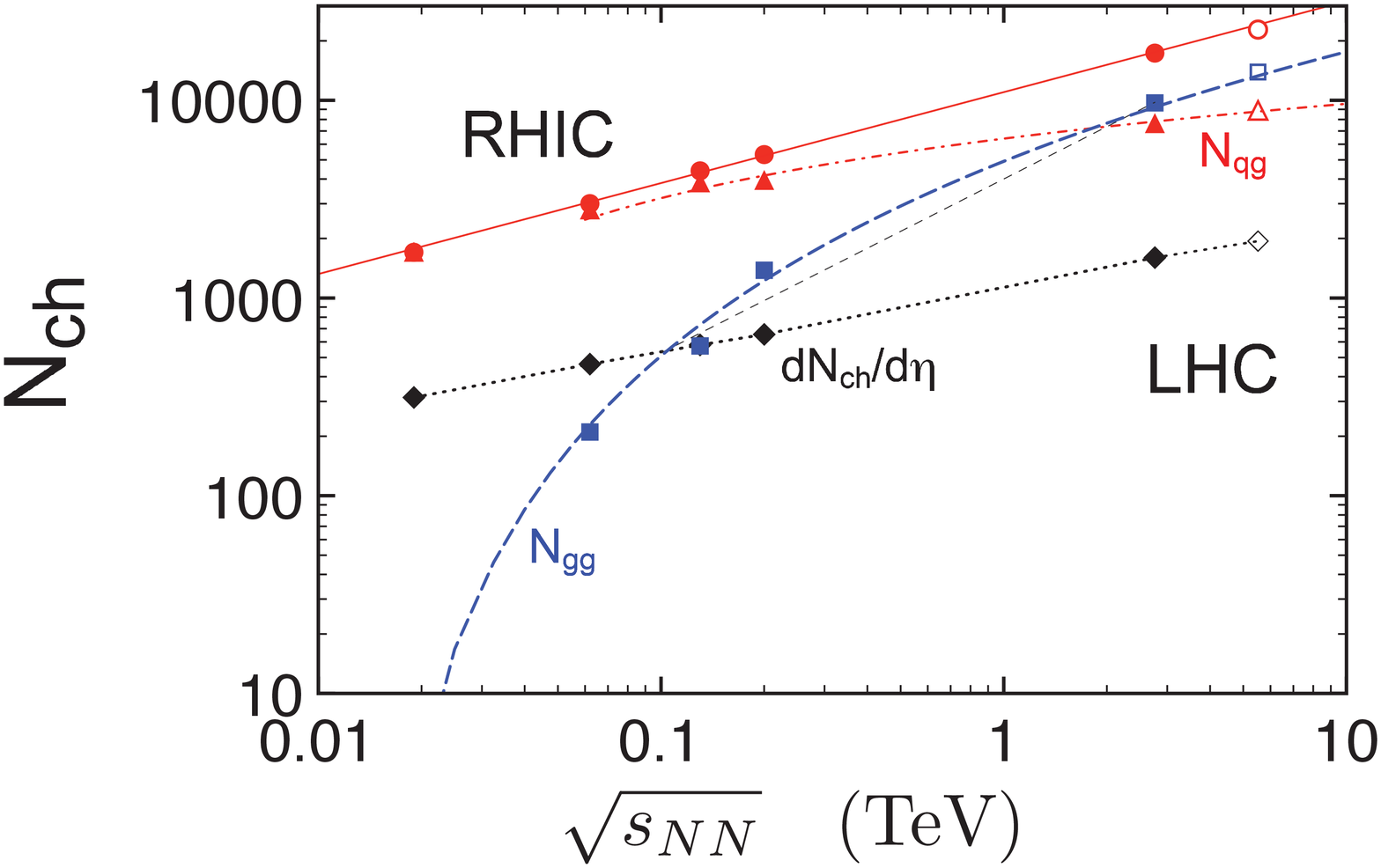}
\caption{\label{fig4}(Color online) The total charged-hadron production in central AuAu and PbPb collision in the energy region 19.6 GeV to 5.52 TeV is following a power law $N_{tot }\propto (s_{NN}/s_0)^{0.23}$ (solid line), whereas the particle content in the fragmentation sources is $Nqg \propto \ln{(s_{NN}/s_0})$, dash-dotted curve. The particle content in the mid-rapidity source obeys $N_{gg} \propto \ln^3{(s_{NN}/s_0)}$,
dashed curve, not too far from a power law (short-dashed line) only in the intermediate energy range 0.1--2.76 TeV. The energy dependence of the mid rapidity yield is shown as a dotted line, with PHOBOS data \cite{alv11} at RHIC energies,
and ALICE data \cite{aamo11} at 2.76 TeV.}
\end{center}
\end{figure}

Results of this approach are summarized in Fig.~\ref{fig1}, where the charged-hadron pseudorapidity distributions are shown from low RHIC energies of 19.6 GeV, via 130 GeV, 200 GeV, to 2.76 TeV, plus a prediction at 5.52 TeV. It is noted that the midrapidity source is found to be absent at 19.6 GeV and appears only at the higher energies, rising in particle content with $\sqrt{s_{NN}}$. The individual sources are displayed in Fig.~\ref{fig2} at 200 GeV and 2.76 TeV, where the effect of the Jacobian transformation from rapidity $y$ to pseudorapidity $\eta$ is also shown. The central gluon-gluon source is seen to become dominant as the energy is increased from RHIC to LHC.

The corresponding particle contents of the sources are displayed in Fig.~\ref{fig3}, which resembles the analogous figure in \cite{gw13}, but differs in a decisive detail. The total particle content is found to follow a power law,
\begin{equation}
N_{ch}^{tot}=1.1\cdot 10^4 (s_{NN}/s_0)^{0.23}
\end{equation}
with  $s_0=1~$TeV$^2$, whereas the particle content in the two fragmentation sources is as expected a logarithmic  function of the energy
\begin{equation}
N_{ch}^{qg}=695\cdot \ln(s_{NN}/s_0)
\end{equation}
with $s_0=100~$GeV$^2$.
The midrapidity gluon-gluon source is approximated by the thin dashed line following a power law as was already proposed in \cite{gw13}
\begin{equation}
N_{ch}^{gg} \simeq4\cdot 10^3(s_{NN}/s_0)^{0.44}
\end{equation}
with $s_0=1~$TeV$^2$. However, when considering also the yield predicted within the relativistic diffusion model (RDM) for the LHC design energy of 5.52 TeV, the power law fails to fit the expected yield, whereas a cubic log dependence agrees with the prediction,
\begin{equation}
N_{ch}^{gg}=7.5\cdot \ln^3(s_{NN}/s_0)
\end{equation}
where $s_0=$169 GeV$^2$.

It remains to be seen whether the data actually follow the model prediction. In the upcoming PbPb run at the LHC in 2015, the c.m. energy is scheduled to be 5.125 TeV, corresponding to 13 TeV $pp$. The total charged-hadron yield predicted by Eq.~(2) at this energy is $N_{ch}^{tot} = 23,327$, with the central source contributing $N_{ch}^{gg} = 12,811$
charged hadrons according to Eq.~(5). The RQM-value for the total charged-hadron production at the lower LHC energy of 2.76 TeV is $N_{ch}^{tot} = 17,327$ according to Tab.~1 of \cite{gw13}; the power law Eq.~(2) yields 17,546. The ALICE collaboration meanwhile quotes an extrapolated value of $17,146\pm 722$ \cite{abb13}. 

When examining the RDM results for the particle content of the sources more closely also in the low-energy region where RHIC data are available, it turns out that the power law 
Eq.~(4) is an acceptable approximation to $N_{ch}^{gg}$ only between about 100 GeV and 2.76 TeV. 

This becomes particularly obvious in Fig.~\ref{fig4}, where the same plot is  shown using a double-logarithmic scale, following a suggestion by Trainor \cite{tr14}. Here power laws appear as straight lines -- such as the one for the total charged-hadron production, or also for the midrapidity yield
\begin{equation}
\frac{dN_{ch}^{tot}}{d\eta}|_{\eta\simeq0}=1.15 \cdot 10^3 (s_{NN}/s_0)^{0.165}
\end{equation}
with $s_0=1~$TeV$^2$ (dotted line, and data points from Phobos \cite{alv11} and ALICE \cite{aamo11}). 

The cubic-log dependence of the gluon-gluon source (dashed) is seen to fit the points extracted from the RDM-analyses \cite{gw13} of PHOBOS and ALICE data rather precisely at the available energies, and it agrees with the RDM-prediction at the LHC design energy of 5.52 TeV. 

As required by the RDM analysis of the 19.6 GeV AuAu data, the gluon-gluon contribution becomes unimportant below 20 GeV -- whereas a power law would still predict a yield of about 100 charged hadrons in this energy region. Although a hybrid function with a log-dependence at RHIC energies that turns into a power law at LHC energies may appear as a reasonable compromise \cite{sami14,sah14}, it can not compete with the ln$^3$-dependence for the central source regarding the precision of the fit to the RDM-results.

\section{Energy dependence of the mid-rapidity source}

The origin of the cubic-log dependence of the total charged-hadron yield on $s_{NN}$ (or 
$\sqrt{s_{NN}}$) in the mid-rapidity gluon-gluon source can be traced schematically 
neglecting for the moment the precise value of the proportionality factor appearing in Eq.~(5).
The width of the gluon-gluon distribution is expected to scale roughly with the beam rapidity,
\begin{equation}
\label{width}
\sigma \propto y_{\text{beam}} =  \ln{(\sqrt{s_{NN}}/m_p)}=0.5\ln{({s_{NN}}/m_p^2)}
\end{equation}
where $m_p$ is the proton mass.
With respect to the midrapidity value, the STAR collaboration observed in 2004-2006
that dijet production Ñ which generates the hard component of the spectrum Ñ is at
midrapidity proportional to the square of the soft-component density,
that is associated with low-$x$ gluons \cite{star06,trai14}. 

Since the yield of low-$x$ gluons is proportional 
to the logarithm of the c.m. energy, the density at midrapidity that arises from
dijet production is proportional to $\ln^2s$. 
Hence, the integrated yield in the gluon-gluon source can be estimated as
\begin{equation}
\label{ngg}
N_{ch}^{gg} \simeq \int_{-y_{\text{beam}}}^{y_{\text{beam}}}\frac{dN}{d\eta}|_{gg}d\eta\propto\ln^3(s_{NN}/s_0)\\ 
\end{equation}
in agreement with the above result of the phenomenological RDM-analysis.

On the theoretical side, the $gg \rightarrow gg$ scattering amplitude has been evaluated in the presence of a classical color field e.g. by Cheung and Chiu \cite{cheu11}.
They find that the classical color field modifies the $gg\rightarrow gg$ elastic scattering amplitude, and suppresses it when the longitudinal momentum fraction $x$ of the incident gluon is small. The rise of the cross section with energy in the central distribution -- that is driven by the
growth of the gluon density at small $x$ -- is therefore suppressed by the quantum-classical interaction from the dense medium \cite{cheu11}.
The predicted cross section has a ln$^2s$ asymyptotic behavior that satisfies the Froissart bound \cite{fro61}, and the integral over
rapidity becomes proportional to $\ln^3s$.





It is interesting to compare the results of the present analysis with the rapidity distributions from the hydrodynamic approach of Landau and Belen'kji  \cite{lan53,lan55}, and applications to particle production by Carruthers and Duong-van \cite{car72,car73}, as well as Steinberg \cite{stei06}. There the width (FWHM) $\Gamma = \sqrt{8\ln2}\cdot\sigma$ of a gaussian pseudorapidity distribution for produced charged particles is obtained from the variance \cite{car73} 
\begin{equation}
\label{landau}
\sigma^2_{\text{Landau}}= \ln \gamma = \ln{(\sqrt{s_{NN}}/2m_p)}
\end{equation}
with the Lorentz-factor $\gamma = 1/\sqrt(1-\beta^2), \beta = p/E.$ 
It turns out that this expression is in reasonable agreement \cite{stei06} with data from AGS and SPS where the stopping fraction is sizeable. Deviations start to become visible at RHIC -- where nuclear transparency \cite{bjo83} with well-separated fragmentation sources is already obvious -- and, in particular, at LHC where the measured width of the $dN/d\eta$-distributions for charged hadrons is substantially broader than predicted by Eq.~(\ref{landau}), as shown by the ALICE collaboration \cite{abb13}. It has therefore been concluded '...that Landau hydrodynamics does not explain the expansion dynamics at LHC energies' \cite{sah14}.

Whereas this is certainly true for the overall pseudorapidity distribution of charged particles, Landau's approach may still be viable for a proper description of the mid-rapidity source
which accounts for particles generated from low-$x$ gluons. Indeed for 2.76 TeV PbPb, the RDM-analysis yields a width in rapidity $y$ of $\Gamma_{gg} = 6.24$ \cite{gw13}, compared to a Landau result of $\Gamma_{gg}^{\text{Landau}} = 6.36$ in $\eta$ according to Eq.~(\ref{landau}). At RHIC energies, the Landau result is, however, larger than the RDM result for the mid-rapidity source, and the results at the higher LHC energies of 5.125 TeV and 5.519 TeV PbPb remain to be seen.

\section{Yields beyond the beam rapidity}
\begin{figure}[tph]
\begin{center}
\includegraphics[width=8.8cm]{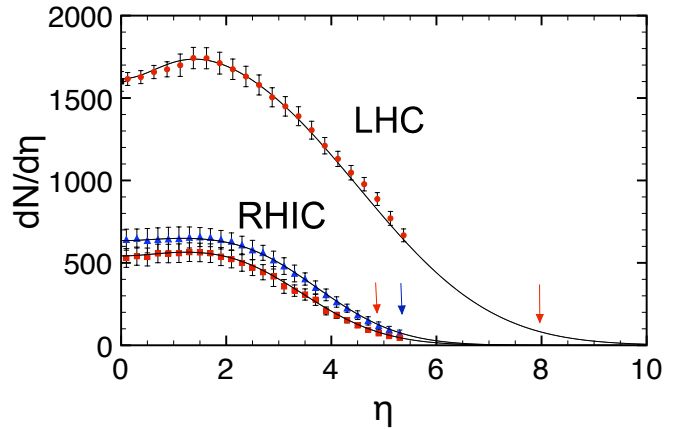}
\caption{\label{fig5}(Color online) Produced charged particles in central AuAu collisions at $\sqrt{s_{NN}} = 130$ and $200$ GeV (RHIC/ PHOBOS data \cite {alv11}, bottom, and in 2.76 TeV PbPb \cite{gui13}, top, in comparison with the RDM solutions. The values of the beam rapidities are indicated as arrows ($y_{\text{beam}}=4.932, 5.362, 7.987)$. The pseudorapidity yields extend beyond $y_{\text{beam}}$, which is particularly evident in case of the 130 GeV PHOBOS data.}
\end{center}
\end{figure}

Already in the investigation of AuAu collisions at RHIC energies \cite{alv11} it has been observed that pseudorapidity yields of produced charged particles extend significantly beyond the value of the beam rapidity. This is particularly obvious in PHOBOS AuAu results at 130 GeV where $dN/d\eta$ data have been taken beyond $y_{\text{beam}}$ \cite{bb01}. The RDM solutions for 200 GeV
AuAu and 2.76 TeV PbPb also clearly indicate expected yields beyond $y_{\text{beam}}$ at these higher energies, see Fig.~\ref{fig5}. 

Obviously it is not excluded that this can partly be due to a real physical effect, with a few charged particles produced at larger rapidities than that of the beam value. However, the bulk of the large charged-particle pseudorapidity density in the region at and beyond the beam rapidity -- which amounts to more than 100 charged particles -- is likely due to the transformation from rapidity to pseudorapidity.

Reconsider the expressions for rapidity $y$, longitudinal velocity $ \beta_{||}$, and pseudorapidity $\eta$
\begin{equation}
y=\frac{1}{2} \ln{\frac{1+\beta_{||}}{1-\beta_{||}}}
\end{equation}
\begin{equation}
 \beta_{||}=\frac{\exp{(2y)}-1}{\exp{(2y)}+1}
\end{equation}
\begin{equation}
\eta=-\ln{(\tan(\theta/2)).}
\end{equation}
\noindent
The transformation between $\eta$ and $y$ is 
\begin{equation}
y=\frac{1}{2}\ln{ \frac{\sqrt{(m/p_T)^2+\cosh^2y}+\sinh\eta}{\sqrt{(m/p_T)^2+\cosh^2y}-\sinh\eta}}. 
\end{equation}
Here $m$ is the mass of the particle species considered. The relative particle abundances in central (0-5\%) PbPb collisions at 2.76 TeV are 
83\% pions, 13\%  kaons, and 4\% protons, with the pion fraction increasing to 84\% for more peripheral (50-60\%) collisions. Hence, I use
an accordingly averaged effective mass for $m$ as described in detail in \cite{rgw12}. 

Since only the ratio $m/p_T$ enters the Jacobian,
one can also fix the mass at the pion mass $m = m_{\pi}$, and calculate the corresponding effective transverse momentum from
$<p_{T,\text{eff}}>=m_{\pi}J_{y=0}/\sqrt{1-J_{y=0}^2} $
with the experimentally determined Jacobian $J_{y=0}$ at $90^0$, see \cite{rgw12} for 2.76 TeV PbPb.
The values of $m/p_T$ used in the calculations shown in Fig.~\ref{fig5} are $m/p_T$ = 0.466; 0.349; 0.585 
for $\sqrt{s_{NN}}$ = 0.13; 0.2; 2.76 TeV, respectively.

The above expression for the transformation from $y$ to $\eta$ has the limits
\noindent
$y\rightarrow\eta-\ln(m/p_T) $ for $m<<p_T$, and
$y\rightarrow \eta$ for $p_T<<m$. Since most of the produced charged hadrons at a LHC energy of 2.76 TeV are pions,
the limit $y \approx \eta$ at small transverse momenta -- very forward angles -- is reached for charged hadrons at larger values of $\eta$ than for protons (net protons determine the value of the beam rapidity). Hence, the $dN/d\eta$ distribution for charged hadrons which are mostly pions can easily extend beyond $y_{\text{beam}}$.

\section{Conclusion}

The analysis of the energy dependence of charged-hadron pseudorapidity distributions in AuAu collisions at RHIC energies, and PbPb collisions at LHC energies in the phenomenological nonequilibrium-statistical relativistic diffusion model reveals the expected ln($s$)-dependence for the total particle content of the two fragmentation sources,
but a ln$^3s$-dependence for the total charged hadron content of the gluon-gluon source. Modifying the conclusion of an initial investigation \cite{gw13}, it is only in a limited energy region of about 100 GeV to 2.76 TeV that this dependence may be approximated by a power law.

\acknowledgments
I thank Tom Trainor for pointing out that a $\ln^3(s_{NN})$-dependence of the gluon-gluon source gives a better representation of
the RDM-results for particle production, than a power law. Discussions with Jean-Paul Blaizot, Larry McLerran and Dmitri Melikhov during their stays at the Institute for Theoretical Physics in Heidelberg are gratefully acknowledged.
\newpage

\bibliography{gw_prc_nt}
\end{document}